\documentclass{ws-procs9x6}

\def\etal {{\it et al.}}

\begin{document}

\title{USING BINARY PULSARS TO TEST LORENTZ SYMMETRY 
IN THE GRAVITATIONAL SECTOR}

\author{J.M.\ WEISBERG}

\address{ Department of Physics and Astronomy, Carleton College\\
Northfield, MN 55057, USA\\
E-mail: jweisber@carleton.edu}

\begin{abstract}
I review some of the major developments in the theoretical background and 
experimental uses of binary pulsars to explore 
local Lorentz invariance in the gravitational sector and its possible violation.\end{abstract}

\bodymatter

\section{Binary pulsars}

A binary pulsar (PSR) consists of a spinning neutron star (NS) with a radio beam, and a companion, in
a mutual orbit.    The companion is usually another compact object --- either
another NS or a white dwarf (WD).   
A typical NS is a  highly 
relativistic object.  With  mass $M \sim 1.4 M_{\rm Sun}$ and radius $R \sim 10$ km,  the
NS has  $G M / ( c^2 R ) \sim 0.2$.     Furthermore,  the PSR's radio pulses serve
as a highly precise
spin-powered  clock, enabling the accurate measurement of  the spin and orbital parameters of
the binary system and the  testing of  relativistic gravitational effects in strong-field
conditions.  For example, the double NS system  PSR B1913+16 has provided
the first evidence for the existence of gravitational waves,\cite{wnt10}  while 
NS-WD systems tend to be the best binary PSRs for probing violations 
of Lorentz invariance. 

\section{PPN extensions and PFE in binary PSRs}

The  weak-field, parametrized post-newtonian (PPN) framework was extended in 
Ref.\ \refcite{wn72} 
to include the possibility of  preferred-frame effects (PFE), via parameters  $\alpha_1,
\alpha_2,  \text{ and }  \alpha_3$.  This analysis was further 
augmented to encompass  the strong fields of binary PSRs,\cite{def92} as follows.   

\begin{figure}
\begin{center}
\psfig{file=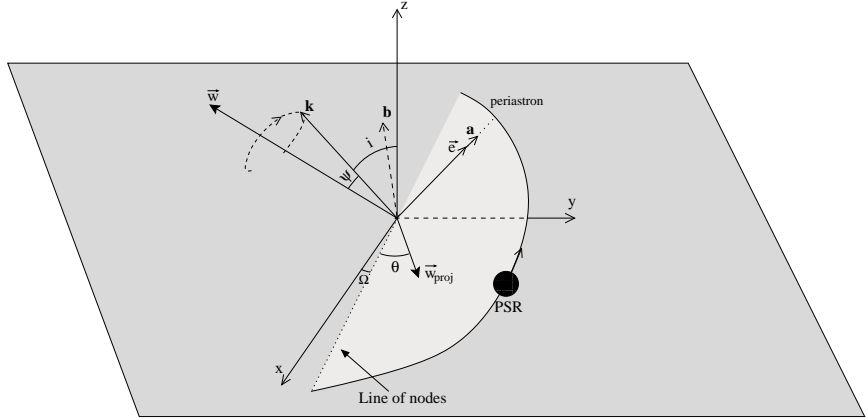,width=\textwidth}
\end{center}
\caption{ PSR orbital geometry.  Unit  vectors  $\left\{   {\textbf a,\textbf  b, \textbf k} \right\}$ define  an  
orthogonal coordinate system 
centered on the binary's center of mass,   with   {\textbf a} and {\textbf b} 
in the orbital  plane (lightly shaded) and {\textbf k}  pointing in the direction of orbital angular 
momentum.  Coordinate {\textbf a} points 
toward PSR periastron, and $ {\textbf a} \times  {\textbf b} = {\textbf k}$.  
The eccentricity $e$ is incorporated into the {\it{vector}} $\vec{\textbf e}= e\  {\textbf a}$.  
The unit vectors 
$\left\{   {\textbf x,\textbf  y, \textbf z} \right\}$ define another orthogonal coordinate system, 
with   {\textbf x} and {\textbf y} in the  
plane of the sky (heavily shaded), and  $ {\textbf z}$  pointing  
{\it{from}} the observer {\it{to}} the binary's center of mass.  
Unit vector $\textbf x$ points toward celestial north.  
This right-handed coordinate system is opposite
to the observers' convention for the plane of the sky.
The velocity of the binary system's center of mass
with respect to the preferred frame  and its projection onto the orbital plane are $\vec{\textbf w}$ 
and $\vec {\mathbf { w}}_{\rm proj}$, respectively.  Adapted from Ref.\ \refcite{sw12}.}
\label{aba:fig1} 
\end{figure}

A gravitational $N$-body post-newtonian lagrangian 
was derived with terms for $\hat{\alpha}_1$ and 
$\hat{\alpha}_2$,
where the `hat' over a  parameter indicates that it is a strong-field modification of
the PPN version; commensurate with  the highly relativistic environment 
of a binary PSR system. 
The $\hat{\alpha}_1$-part of the lagrangian includes a dependence of the vector 
velocity $\vec{\textbf w}$ of 
the  binary center of mass with respect to the preferred frame, 
leading to some potentially measurable effects.
For small-eccentricity binary PSRs,    the eccentricity vector 
$\vec{\textbf e} \equiv e\  {\textbf a}$ (see Fig.\ 1) undergoes the following time evolution:
\begin{equation}
\vec{\textbf e}(t)=\vec{\textbf e}_{\rm Rel}(t) + \vec{\textbf e}_{\rm Fixed};
\label{aba:rotecc}
\end{equation}
where the first term results from the usual general relativistic  advance of periastron whose mean 
rate is  $\langle\dot{\omega}_{Rel}\rangle$; but the second is a nonzero constant `polarizing'' term 
resulting from PFE, and 
proportional to $G_{\rm orb}$ (a function of orbital parameters and component masses),
$\hat{\alpha}_1$ and $\vec {\mathbf { w}}_{\rm proj}$ (the projection of  $\vec{\textbf w}$ 
onto the orbital plane):
\begin{equation}
\vec{\textbf e}_{\rm Fixed}=G_{\rm orb} \hat{\alpha}_1  \textbf{k}      \times \vec{\textbf w}_{\rm proj},
\label{aba:fixedecc}
\end{equation}
(which points  in the orbital plane, perpendicular to $ \vec{\textbf w}_{\rm proj}$).
Hence the measurement of eccentricity over time affords the possibility of separating the two terms
in Eq.\ \eqref{aba:rotecc} and then, for a given choice of  preferred frame of $\vec {\mathbf { w}}$ 
(e.g., the cosmic microwave background), 
determining the desired PFE parameter $\hat{\alpha}_1$ from Eq. \ref{aba:fixedecc}.

However, some observational difficulties remain.  First, the velocity of the binary system's
center of mass with respect to Earth is difficult to measure,
which limits the accuracy of $\vec {\mathbf { w}}$ (both its direction and magnitude).
Also the angle $\Omega$ is not measurable from PSR timing, thereby further limiting the 
accuracy of its projection $\vec {\mathbf { w}}_{\rm proj}$. In general, statistical arguments must
be employed in order to ameliorate these limitations, although much 
progress has also been made in overcoming them observationally. (See below.)
The measured eccentricity for the NS-WD PSR B1855+09 system 
was then combined with the above analysis to provide 
an upper limit on $|\hat{\alpha}_1|$ of $5.0 \times 10^{-4}$, which was
comparable to contemporaneous solar system limits,
and (unlike the solar system case) measured in {\it{strong}} gravitational  fields.

Building upon the work of Ref.\ \refcite{def92}, it was shown\cite{sw12} that 
{\it{both}} $\hat{\alpha}_1$ {\it{and}} $\hat{\alpha}_2$ are determinable from timing measurements
of low-eccentricity binary PSRs, and the new analysis was applied to two additional  NS-WD 
binary PSR systems: PSR J1012+5307
and PSR J1738+0333,
as follows.

A nonzero $\hat{\alpha}_2$ will cause precession of the orbital pole, $\mathbf k$, about
$\vec {\mathbf { w}}$
(see Fig.\ 1). This precession leads to
an observable consequence in the  PSR timing measurable 
$x \equiv a_{\rm PSR} \sin i$.  Here $a_{\rm PSR}$ is the semimajor axis
of the PSR orbit, and $i$ is the `inclination' angle between  $\mathbf k$ and  $\mathbf z$:
\begin{equation} 
\left(\frac{\dot{x}}{x}\right)_{\hat{\alpha}_2}=-\frac{\hat{\alpha}_2} {4} 
\left(\frac{2 \pi}{P_{\rm b}}\right)   \left(\frac{w}{c}\right)^2 \cot i \sin 2 \psi  \cos \theta,
\label{aba:xdot}
\end{equation}
where  angles $\psi$ and $\theta$ specify the instantaneous direction of $\vec{\mathbf {w}}$
in the (precessing) $\left\{   {\textbf a,\textbf  b, \textbf k} \right\}$ coordinate system,
as shown in Fig.\ 1. However, there are observational difficulties in determining $w$ and the
angles, similar to those discussed above in the determination of $\hat{\alpha}_1$.  While significant
progress has been made, the nature of PSR timing observations  prevents the  determination
of $\Omega$ and hence accurate values of  $w, \theta$, and $\psi$, so it remains necessary
to use statistical arguments to extract the PFE parameter ${\hat{\alpha}_2}. $ 
A joint analysis  from both pulsars leads to $|{\hat{\alpha}_2}| < 1.8 \times 10^{-4}$.
This limit is far weaker than the solar 
system limit of Ref.\ \refcite{n87}, but does have the benefit of probing strong fields.  The
analysis of $|{\hat{\alpha}_1}| $ of Ref.\ \refcite{def92} was also revisited. With the benefit of 
additional kinds of observational data that leave only $\Omega$ undetermined,
an upper limit on ${\hat{\alpha}_1} $  of $(-0.4\  [+3.7, -3.1]) \times 10^{-5}$  was found for PSR 
J1738+0333, making it the best extant constraint for this parameter.  

The determination of PPN parameter $\alpha_3$ from binary PSRs    was derived 
theoretically in Ref.\ \refcite{bd96}; Ref.\ \refcite{G11} reports an exquisitely precise  
null measurement, as expected in semiconservative theories of gravitation.

\section{SME coefficicents for Lorentz violation in binary PSRs}

The first extensive theoretical SME analysis of the  signal of Lorentz violation  in binary PSRs 
was published in Ref.\ \refcite{bk06}. In developing the general theory, 
twenty independent Lorentz-violating coefficients were 
found, including nine in the field $\bar{s}^{\mu \nu}$.  It was shown that three of the
PPN parameters are   expressible
in terms of the single  coefficient $\bar{s}^{0 0}$; and 
that $\alpha_1=4 \alpha_2$, indicating that (at least in the weak-field limit), they are not 
independent.  More importantly, it was  shown that binary PSR  secular  {\it{measurables}} 
$\langle\dot{\omega}_{Rel}\rangle, \langle de / dt\rangle,$ and $\langle di/dt \rangle$ 
can be expressed in  terms of 
combinations of projections of the SME $\bar{s}^{\mu \nu}$ along the orbital axes 
$\left\{   {\textbf a,\textbf  b, \textbf k} \right\}$, raising the prospect of determining these SME
Lorentz-violating coefficients via binary PSR timing observations.  Therefore wholly new
tests of Lorentz violation become available which are not accessible in the PPN formulation.

\section{Future prospects}

Observationally, the most promising direction involves the detection and measurement of additional
binary PSR systems.  These would increase the parameter-space coverage of 
 current tests, 
which could lead for example to disentangling the currently covariant SME Lorentz-violation 
coefficients\cite{bk06}, and increasing the sky coverage of Lorentz-violation tests.\cite{sw12}
Planned new, highly sensitive radiotelescopes will detect numerous such systems\cite{c04}.  Theoretically, 
the most exciting  prospect includes
the  generation of additional  SME measurables in binary PSR systems.

\section*{Acknowledgments}
This research was supported by NSF grant AST-0807556.


\begin{thebibliography}{x}

\bibitem{wnt10}  
J.M.\ Weisberg, D.J.\ Nice, and J.H.\ Taylor, 
Ap.\ J.\ {\bf{722}}, 1030 (2010).

\bibitem{wn72}  
C.M.\ Will and K.\ Nordtvedt Jr., 
Ap.\ J.\ {\bf{177}}, 757 (1972).

\bibitem{def92} 
T.\ Damour and G.\ Esposito-Far{\`e}se, 
Phys.\ Rev.\ D {\bf{46}}, 4128 (1992).

\bibitem{sw12}
L.\ Shao and N.\ Wex, 
Class.\ Quantum Grav.\ {\bf{29}}, 215018 (2012).

\bibitem{n87} 
K.\ Nordtvedt, 
Ap.\ J.\ {\bf{320}}, 871 (1987).

\bibitem{bd96}  
J.F.\ Bell and T.\ Damour,  
Class.\ Quantum Grav.\ {\bf{13}}, 3121 (1996).

\bibitem{G11} 
M.E.\ Gonzalez \etal, 
Ap.\ J.\ {\bf{743}}, 102 (2011).

\bibitem{bk06} 
Q.G.\ Bailey and V.A.\ Kosteleck\'y,
Phys.\ Rev.\ D {\bf{74}},  045001 (2006).

\bibitem{c04} 
J.M.\ Cordes \etal,
New Astron.\ Rev.\ {\bf{48}}, 1413 (2004).

\end{thebibliography}
\end{document}